\documentclass[10pt]{article} 

\usepackage{geometry}
\usepackage[
	dvips,
	colorlinks,
	citecolor = blue,
	linkcolor = blue,
	urlcolor = blue,
	]{hyperref}

\usepackage{fancyhdr}
\usepackage{graphicx}
\usepackage{booktabs}
\usepackage{multirow}

\usepackage{theorem}
\newtheorem{definition}{Definition}[section]

\usepackage{caption,subcaption}
\usepackage{hyperref}
\usepackage{lineno}
\usepackage{mathtools,txfonts,upgreek}

\usepackage{algorithm}
\usepackage{algpseudocode}
\usepackage[section]{placeins}
\usepackage{url}

\graphicspath{{./Figures/}}

\title{Tri-criterion model for constructing low-carbon mutual fund portfolios: a preference-based multi-objective genetic algorithm approach}

\usepackage{authblk}
\author[1]{A. Hilario-Caballero}
\author[1]{A. Garcia-Bernabeu}
\author[1]{J. V. Salcedo}
\author[1]{M. Vercher}
\affil[1]{Universitat Politècnica de València}

\providecommand{\keywords}[1]{
	\vspace{2ex}
	\small	
	\noindent\textbf{\textit{Keywords---}} #1
	}

\begin{document}

\maketitle

\begin{abstract}
Sustainable finance, which integrates environmental, social and governance (ESG) criteria on financial decisions rests on the fact that money should be used for good purposes. Thus, the financial sector is also expected to play a more important role to decarbonise the global economy. To align financial flows with a pathway towards a low-carbon economy, investors should be able to integrate in their financial decisions additional criteria beyond return and risk to manage climate risk.  We propose a tri-criterion portfolio selection model to extend the classical Markowitz mean-variance approach in order to include investors preferences on the portfolio carbon risk exposure as an additional criterion. To approximate the 3D Pareto front we apply an efficient multi-objective genetic algorithm called ev-MOGA which is based on the concept of $\epsilon$-dominance. Furthermore, we introduce an a posteriori approach to incorporate the investor's preferences into the solution process regarding their sustainability preferences measured by the carbon risk exposure and his/her loss-adverse attitude. We test the performance of the proposed algorithm  in a cross section of European SRI open-end funds to assess the extent to which climate related risk could be embedded in the portfolio according to the investor's preferences.


\end{abstract}

\section{Introduction}

Climate change will pose a challenge for the financial sector seeking a balance between purely financial goals -- looking for high returns -- and sustainability making a positive impact on the environment and on society. Since 2015, by adopting the Paris Agreement on climate change and the UN 2030 for Sustainable Development, there has been a clear commitment, especially in the European Union, to align financial flows with a pathway towards a low-carbon, more resource-efficient and sustainable economy. In 2018, the EU has launched and Action Plan to set out a strategy for sustainable finance, that is , ``the process of taking due account of environmental and social considerations in investment decision-making, leading to increased investments in longer-term and sustainable activities'' \cite{EU2018}. As stated in this report, to date, environmental and climate risks had not been appropriately considered by the  financial sector, which is why if the EU wants to reorient capital flows to a more sustainable economy, environmental and social goals will have to be included in the financial decision-making. To this end, the \textit{Markets in Financial Instruments Directive (MIFID II)} and the \textit{Insurance Distribution Directive (IDD)} provide that investment firms and insurance distributors should ask their clients' investments objectives as regard sustainability and take their preferences into account when providing financial advice. This implies that investors should be able to integrate in their financial decisions additional criteria beyond return and risk and then to extend the classical bi-criterion portfolio selection problem based on Markowitz mean-variance approach \cite{markowitz1952portfolio} by adding one more criterion.

In the literature, tri-criterion portfolio selection problems have been addressed by several authors making use of multicriteria decision problems (MCDM). One of the first attempts to compute the variance-expected return-sustainability surface was \cite{hirschberger2013computing}. These authors, proposed an inverse portfolio optimization algorithm using CIOS (Custom Investment Objective Solver) from the model of Markowitz and they generated a tri-criterion non dominanted surface composed of  a connected collection of parabolic "platelets''.  In \cite{utz2014tri} and \cite{utz2015tri}, the previous procedure was applied to construct a model that includes risk, expected return and sustainability, which is measured using ESG scores.  In recent years, Multi-objective Evolutionary Algorithms (MOEAs) have been proposed to handle two or more conflicting goals subject to several constraints \cite{arnone1993genetic} and in particular to address complex portfolio selection problems \cite{metaxiotis2012multiobjective}. A recent approach based on ev-MOGA \cite{herrero06ingles} has been adapted in \cite{garcia2019computing} to derive the non-dominated mean-variance-sustainability surface. 

In line with the goals of the Paris Agreement, the financial flows should be consistent with a pathway towards
low greenhouse gas emissions. In recent years, in response to an increasing  climate-conscious financial products demand, Morningstar,
the most important information provider in the mutual fund industry, introduced the Low Carbon Designation eco-label \cite{Morningstar_2018}. This new mutual fund eco-label helps investors to easily recognize which mutual funds are aligned with the transition to a low-carbon economy \cite{Ceccarelli2020}. The LCD is composed of two indices, the Carbon risk score and the Fossil Fuel involvement. In our research we only consider the fund-level Carbon Risk score (from ESG Sustainalytics provider) which is obtained by weighting the firm-level exposure and management of material carbon issues. As \cite{krueger2020importance} highlights, institutional investors increasingly address climate related risk and they are also viewed as catalysing driving firms to meet the reduced emission target. Thus, the  mutual fund industry, and in particular  institutional investors is an ideal setting to test our proposal.

Our paper makes the following contributions to the literature. First, it gives a better understanding of recent multi-criteria decision making methodologies (MCDM) to deal with tri-criterion portfolio selection problems by reviewing the literature on exact methods and multi-objective genetic algorithms techniques. Second, it allows us to integrate carbon risk exposure as a new objective in the portfolio optimization procedure of institutional mutual funds. We then,  propose a recent multi-objective genetic algorithm called ev-MOGA \cite{herrero06ingles} to provide investors with the insights to make more informed decisions and to manage portfolio carbon risk expose more effectively. Third, the preferences of the decision maker are incorporated  into the solution process regarding their climate or green preferences measured by the carbon risk exposure and their loss-adverse attitude measured by the variance of returns. Taking into account the green preferences, we define three investor profiles: weak green investor, moderate green investor and strong green investor. Moreover, we also consider their attitude towards risk and then, we define three types of profiles: conservative, cautious and aggressive.

The rest of the paper is structured as follows. Section 2 starts with a review of the literature about the tri-criterion portfolio selection problem addressed either by exact or heuristic methodologies. In Section 3, the proposed tri-criterion genetic multi-objective evolutionary algorithm  for constructing low carbon portfolios is formulated including the a-posteriori approach to integrate sustainability preferences in financial decision-making.  In Section 4, we analyze the numerical results obtained by the application of the ev-MOGA using different investor profiles for a data set of  European Socially Responsible Investments (SRI) open-end funds. Finally, Section 5 concludes the paper.


\section{Exact methods vs Multi-objective Evolutionary Algorithms for extended M-V portfolio selection: a literature review}

The idea of determining the Pareto efficient frontier in portfolio selection from a mean-variance (M-V) optimization was originally conceived in \cite{markowitz1952portfolio}. The essence of the  M-V model is that risk is the investor's main concern and he/she tries to minimize risk for a desired level of expected returns. Over the years, the Markowitz model  has been extended either through more complex risk measures or through additional constraints, and in recent years through the possibility to include additional objectives. In this context,  two main approaches to deal with the extended portfolio optimization problem can be found: (i) Exact methods or (ii) Heuristic methodologies. 

\subsection{M-V extended approaches by exact methods}

Since the early 1970s several authors have attempted to expand the classical bi-criteria portfolio selection model beyond the expected return and variance with exact methods. Three main groups of studies can be identified dealing with this problem. A first group of authors have expanded the Markowitz model by introducing additional constraints such as cardinality, round lots or  buy-in threshold \cite{syam1998dual,mansini2002exact,li2006optimal, bertsimas2009algorithm}.  Alternative risk measures such as down-side risk measures or CVaR have been proposed in a second group of studies \cite{bawa1975optimal,konno1991mean,rockafellar2001conditional}. A literature review on risk measures in terms of computational comparison is conducted in \cite{mansini2003lp}. 

 Not until the 20th century was the idea of additional objectives was further boosted by a third group of studies. A tri-criterion non dominated surface can be found in \cite{hirschberger2013computing,utz2014tri,utz2015tri} using a constrained linear program (QCLP) approach by solving a quad-lin-lin optimization problem where the third objective is linear. By defining several measures of liquidity in \cite{lo2003-liquidity} a three-dimensional mean-variance-liquidity frontier is constructed.  A general framework for computing the non-dominated surface in tri-criterion portfolio selection that extends the Markowitz portfolio selection approach to an additional linear criterion (dividends, liquidity or sustainability) is addressed in \cite{hirschberger2013computing}. By solving a quad-lin-lin program, they provide an exact method for computing the non-dominated surface that can outperform standard portfolio strategies for multicriteria decision makers. An empirical application where the third criterion is sustainability is developed to illustrate how to compose the non-dominated surface.

In \cite{utz2014tri} sustainability is included as the third criterion to obtain the variance-expected return-sustainability efficient frontier in order to explain how the sustainable mutual fund industry can increase its levels of sustainability. The tri-criterion non-dominated surface is computed through the Quadratic Constrained Linear Program (QCLP) approach, and from the experimental results it can be concluded that there was room to expand the sustainability levels without hampering the levels of risk and return. 

However, the existing proposals based on exact procedures to solve tri-criteria portfolio selection problems have limited capabilities when the third objective is non-linear. In such cases, heuristic techniques have been recently applied to solve multi-objective problems and to provide fair approximations of the pareto front.

\subsection{MOEAs and the extended M-V portfolio optimization problem } \label{s:MOEAS_portfolio}

The increasing complexity of financial decision making problems has led researchers to apply heuristic procedures inspired by biological processes such as Multi-objective Evolutionary Algorithms (MOEAs). Suggested in the beginning of the 90s, MOEAs have been applied in several fields including finance,  and in particular to solve the portfolio selection problem \cite{maringer2006portfolio,coello2006evolutionary}.  These techniques provide satisfactory approximations of the efficient frontier even when the problem involves non-convexity, discontinuity or non integer variables. In \cite{arnone1993genetic}, a MOEA was proposed for the first time for optimal portfolio selection by using lower partial moments as a measure of risk. The first attempts to propose MOEAs as an extension of the M-V model aimed at considering additional constrains such as, cardinality, lower and upper bounds, transaction costs, transaction round lots, non-negativity constraints or sector capitalization constraints \cite{chang2000heuristics, maringer2003optimization,shaw2008lagrangian,bertsimas2009algorithm,soleimani2009markowitz,deng2010swarm,anagnostopoulos2011mean,woodside2011heuristic,meghwani2017multi,liagkouras2018new}. A review of the state of the art of MOEAs in portfolio selection can be found in \cite{metaxiotis2012multiobjective}.

Another group of researchers have also tried to propose alternative risk measures to variance, the most popular being: semivariance, value at risk (VaR) and conditional value at risk (CVaR), the lower partial moments (LPM), the Expected Shortfall, the Skewness, and Risk parity \cite{gilli2006data,chang2009portfolio,hochreiter2007evolutionary, liagkouras2019new,kaucic2019portfolio}

Regarding the number of objectives, while the two-objective case is the most widely used among the authors, the  tri-objective problem has risen in popularity in the last few years. A tri-objective optimization problem is proposed in \cite{anagnostopoulos2010portfolio}  to find the trade-off between risk, return and the number of securities in the portfolio. In this paper, the authors compare three evolutionary multi-objective optimization techniques for finding the best trade-off between risk, return and the cardinality of the portfolio.  A recent approach based on ev-MOGA \cite{herrero06ingles} has been adapted in \cite{garcia2019computing} to derive the non-dominated mean-variance-sustainability surface. 

\section{The tri-criterion multi-objective approach by ev-MOGA to manage carbon risk exposure}

During the last two decades, MOEAs for portfolio management have attracted scholars and practitioners attention as stated in \autoref{s:MOEAS_portfolio}. Next, some previous notions on multi-objective optimization and genetic multi-objective optimization techniques are provided.

\subsection{Background on multi-objective optimization and ev-MOGA}


\newcommand{\p}{\omega}
\newcommand{\vp}{\mathbf{w}}
\newcommand{\Vp}{\mathbf{\Upomega}}

\newcommand{\ParetoFrontParameters}{\Vp_{\!P}}
\newcommand{\ParetoFront}{\mathbf{f}(\Vp_{\!P})}
\newcommand{\epsilonParetoFrontParameters}{\hat{\Vp}^*_{\!P}} 
\newcommand{\epsilonParetoFront}{\mathbf{f}(\hat{\Vp}^*_{\!P})}
\newcommand{\epsilonPARETOFRONT}{\mathbf{f}(\hat{\Vp}_{\!P})}



Multi-objective optimization is an important subclass of multiple criteria decision making techniques involving more than one objective function to be optimized simultaneously. Since the conflict degree between the objectives makes it impossible  to find a feasible solution that simultaneously optimizes all the objective functions, there is a set of Pareto optimal solutions denoted as Pareto front at which none of the objectives can be improved without deteriorating at least one of the others. In general a MOP optimization problem is stated as follows:
\begin{equation}\label{eq:MOP}
	\begin{split}
		&\underset{\vp}{\text{minimize}} \qquad \mathbf{f}(\vp) = \left[\,f_1(\vp),f_2(\vp),\ldots, f_m(\vp)\,\right]^T,\\
		&  \text{subject to} \qquad \vp\in S,
	\end{split}
\end{equation}


\noindent where the vector $\vp=[\p_1,\p_2,\ldots,\p_n]^T$ {is a $n$-parameter set included in the decision space} $S$, and $f_i(\vp): \varmathbb{R}^n\rightarrow\varmathbb{R}$, $i=1,\ldots,m$, are the objectives to be minimized at the same time. 

In recent years MOEAs have been widely accepted as useful tools for solving real world multi-objective problems. Within MOEAs several powerful stochastic search techniques that mimic Darwinian principles of natural selection are included. In this study we focus on the ev-MOGA algorithm proposed in \cite{herrero06ingles}, which combines the concept of Pareto optimality  and  $\epsilon$-dominance due to \cite{laumanns2002combining}, thus providing an approximated $\epsilon$-Pareto set. 

\begin{definition}
Dominance: Let $\vp^1,\,\vp^2 \in \varmathbb{R}^n$ be two feasible solutions, an let $\mathbf{f}(\vp^1),\,\mathbf{f}(\vp^2) \in \varmathbb{R}^m$ be their image solutions in the objective space. Then, assuming that the objective functions have to be minimized, $\vp^1$ is said to dominate $\vp^2$, denoted as $\mathbf{f}({\vp}^1)\prec\mathbf{f}({\vp}^2)$, iff:
\begin{equation}\label{eq:dominance}
\begin{split}
	&\forall i \in  \{1,\ldots, m\} \, : \, f_i(\vp^1) \leq f_i(\vp^2)\\
	& \exists\, j \in \{1,\ldots, m\} \, : \, f_j(\vp^1) < f_j(\vp^2)
\end{split}
\end{equation}
\end{definition}


\begin{definition}
Pareto set or Pareto front: Let $\Omega \subseteq \varmathbb{R}^{n}$ be a set of vectors of feasible solutions with $\mathbf{f}(\Omega)$ as their image solutions. Then the Pareto set $\ParetoFront$ of $\mathbf{f}(\Omega)$ is defined as follows: $\ParetoFront$ contains all vectors $\mathbf{f}(\vp^u) \in \mathbf{f}(\Omega)$ that are not dominated by any vector $\mathbf{f}(\vp^v) \in \mathbf{f}(\Omega)$
, i.e.,
\begin{equation}
	\ParetoFront := \left\{\,
		\mathbf{f}(\vp^u) \in \mathbf{f}(\Omega) \;\;|\;\; \nexists\,\mathbf{f}(\vp^v) \,:\,
		\mathbf{f}(\vp^v) \prec \mathbf{f}(\vp^u)
		\,\right\}
\end{equation}
\end{definition}


\begin{definition}
$\epsilon$--dominance: Let $\vp^1,\,\vp^2 \in \varmathbb{R}^n$ be two feasible solutions, an let $\mathbf{f}(\vp^1),\,\mathbf{f}(\vp^2) \in \varmathbb{R}_+^m$ be their image solutions in the objective space. Then $\vp^1$ is said to $\epsilon$--dominate $\vp^2$ for some
$\epsilon > 0$, denoted as $\mathbf{f}({\vp}^1)\prec_\epsilon\mathbf{f}({\vp}^2)$, iff:
\begin{equation}
	\forall\,i \in \{1,\ldots, m\} \,:\, (1+\epsilon)\cdot f_i(\vp^1)\leq f_i(\vp^2)
\end{equation}
\end{definition}


\begin{definition}
$\epsilon$--approximate Pareto set: Let $\Omega \subseteq \varmathbb{R}^{n}$ be a set of feasible solution vectors with $\mathbf{f}(\Omega)$ as their image solutions. Then, $\epsilonParetoFront$ is called a $\epsilon$--approximate Pareto set of $\mathbf{f}(\Omega)$ if any vector $\mathbf{f}(\vp^u) \in \mathbf{f}(\Omega)$ is $\epsilon$--dominated by at least one vector $\mathbf{f}(\vp^v) \in \epsilonParetoFront$
, i.e.,
\begin{equation}
	\forall\,\mathbf{f}(\vp^u) \in \mathbf{f}(\Omega): \exists\,\mathbf{f}(\vp^v) \in \epsilonParetoFront  \;\;|\;\; \mathbf{f}(\vp^v) \prec_\epsilon \mathbf{f}(\vp^u)
\end{equation}
\noindent The set of all $\epsilon$--approximate Pareto sets of $\mathbf{f}(\Omega)$ is denoted as the  $\epsilon$--Pareto front $\epsilonPARETOFRONT$.
\end{definition}



%
%
%
%


The most outstanding feature of this algorithm is that the optimal solutions are distributed uniformly across the $\epsilon$-Pareto front. To this end, the $\epsilon$-Pareto front is split into a fixed number of boxes forming a grid, so that the algorithm ensures that just one solution is stored by one box. The size of the boxes is determined by the value of $\epsilon_i$, which is calculated as follows:
\begin{equation}
\epsilon_i=\frac{{f_i}^*- {f_{i}}_*}{n_{box}}
\end{equation}

\noindent where, $f_i^*$ and $f_{i*}$ correspond to the maximum and minimum value of the objective function $f_i$, and $n_{box}$ is the number of boxes. In addition, ev-MOGA is able to adjust the width of $\epsilon_i$ dynamically and prevent solutions belonging to the extremes of the front from being lost.    

For solving the ev-MOGA, the main population $P(t)$ whose size is $Nind_p$ explores the searching space $S$ defined by the multi-objective problem during a number $k$ of iterations. In the archive population $A(t)$ the $\epsilon_i$-nondominated solutions are stored, so that there are as many feasible solutions as number of boxes. Then, at the end of the iteration process, $A(t)$ is an $\epsilon$-approximate Pareto set $\epsilonParetoFront$.  
Furthermore, in the case that more than one  $\epsilon$-dominant solution is detected, thus the solution that prevails in $A(t)$ will be the one that is closest to the center of the box. Next, the new individuals obtained by crossover or mutation with probability of crossing/mutation $P_{c/m}$ are included in the auxiliary population $G\!A(t)$. 

Before running the algorithm, the following parameters should be defined by the analyst: 

\vspace*{-1ex}

\begin{itemize}\itemsep=0ex
\item $Nind_p=$ Size of the main population.
\item $Nind_{G\!A}=$ Size of the auxiliary population.
\item $k_{max}=$ Maximum algorithm iterations.
\item $P_{c/m}=$ Probability of crossing/mutation.
\item $n_{box}=$ Number of boxes.
\end{itemize}

The main advantage of ev-MOGA is that they generate good approximations of a well-distributed Pareto front in a single run and within limited computational time. The original ev-MOGA algorithm is avalaible at Matlab Central  \cite{ev_MOGA_Matlab_Central_2017}: \href{https://es.mathworks.com/matlabcentral/fileexchange/31080-ev-moga-multiobjective-evolutionary-algorithm}{ev-MOGA in Matlab Central}.

\subsection{The ev-MOGA tri-criterion portfolio selection}

In this study, beyond risk and return, we wish to consider an additional objective that minimizes the carbon risk exposure of a portfolio. Then, by introducing a third objective into the portfolio optimization model the efficient frontier becomes a surface in the three-dimensional space. The tri-criterion portfolio selection problem where the objectives are the risk of the portfolio, the returns, and the portfolio carbon risk exposure can be mathematically formulated as follows:
{\jot=1.25ex
\begin{align}
	\label{eq:tri_01} \min f_1 (\vp) = & \sum_{i=1}^{N}\sum_{j=1}^{N}\p_i\,\p_j\,\sigma_{ij}\\
	\max f_2 (\vp) = & \sum_{i=1}^{N}\p_i\,\mu_i\\
	\label{eq:tri_03} \min f_3 (\vp) = & \sum_{i=1}^{N}\p_i\,c_i\\
	\label{eq:tri_04} \text{subject to} \quad & \sum_{i=1}^{N}\p_i=1
\end{align}
}

\noindent where  $N$ denotes the available assets, $\mu_i$ is the expected return of asset $i$ ($i=1,2,\ldots,N$), $\sigma_{ij}$ is the covariance between asset $i$ and $j$. In addition $c_i$ is the carbon risk score and $\p_i$ denotes the proportion of asset $i$ in the portfolio. 

\makeatletter
\let\OldStatex\Statex
\renewcommand{\Statex}[1][3]{%
  \setlength\@tempdima{\algorithmicindent}%
  \OldStatex\hskip\dimexpr#1\@tempdima\relax}
\makeatother

\begin{algorithm}[ht!]
	\caption{Tri-criterion ev-MOGA algorithm based on \cite{herrero06ingles} }\label{alg:evMOGA}
	\begin{algorithmic}[1]\parskip=0ex
		\Statex
		\State Set $k=0$.  
		\State Initialize  the population of candidate solutions $P_0$ and set $A_0=\emptyset$
		\State Conduct the multi-objective evaluation of portfolios from $P_0$ using Equations \eqref{eq:tri_01}--\eqref{eq:tri_04}
		\State Detect the $\epsilon$-nondominated portfolios from $P_0$ and store in the archive $A_0$
		\While{$k\le k_{max}$}
			\State Generate the auxiliary population $G\!A_k$  from the main population $P_k$ and the
			{\parskip = 1ex
			\Statex[2] archive population $A_k$ following this procedure:}
			\For{$j \gets 1 ,Nind_{G\!A}/2$}
				\State Randomly select two portfolios $\mathbf{X}^{P}$ and $\mathbf{X}^{A}$ from $P_k$ and $A_k$, respectively
				\State Generate a random number $u \in [0,1]$ 
				\State  If $u>P_{c/m}$, $\mathbf{X}^{P}$ and $\mathbf{X}^{A}$ are crossed over by means of the extended linear recombination 
				{\parskip = 1ex
				\Statex[3] technique, generating two new portfolios for $G\!A_k$}
				\State If $u\leq P_{c/m}$, $\mathbf{X}^{P}$ and $\mathbf{X}^{A}$ are mutated using random mutation with Gaussian distribution 
				{\parskip = 1ex
				\Statex[3] and then included in $G\!A_k$}
			\EndFor
		
			\State Evaluate population $G\!A_k$ using  the tri-criterion multi-objective portfolio model defined by  \eqref{eq:tri_01}--\eqref{eq:tri_04}.
			
			\State Check which portfolios in $G\!A_k$ must be included in $A_{k+1}$ on the basis of their location
			{\parskip = 1ex
			\Statex[2] in the objective space. $A_{k+1}$ will contain all the portfolios from $A_k$ that are \Statex[2] not $\epsilon$-dominated by elements of $G\!A_k$, and all the portfolios from $G\!A_k$ which are  
			\Statex[2] not $\epsilon$-dominated by elements of $A_k$}
			
			\State Update population $P_{k+1}$ with portfolios from $G\!A_k$. Every portfolio $\mathbf{X}^{G\!A}$ from $G\!A_k$ is compared 
			{\parskip = 1ex
			\Statex[2] with a portfolio  $\mathbf{X}^P$ that is randomly selected from the portfolios in $P_k$.  $\mathbf{X}^{G\!A}$ will replace 
			\Statex[2] $\mathbf{X}^P$ in $P_{k+1}$ if it dominates $\mathbf{X}^{P}$. Otherwise $\mathbf{X}^P$ will not be replaced}
			\State $k \leftarrow k +1$			
		\EndWhile
		\Statex		
	\end{algorithmic}
\end{algorithm}

\subsection{Defining a-posteriori preferences for each investor's profile}

With the previous multi-objective optimization design a vast region of the tri-objective whole Pareto front is generated. Even though it is true that the non-dominated surface allows us to better understand the trade-off between the three objectives, this solution doesn't provide a useful tool from the user's perspective. To come up with a single solution we assume that the decision maker is available to take part in the solution process. According to \cite{coello2000handling,branke2005integrating} the articulation of preferences may be done either before (a priori), during (progressive), or after (a posteriori) the optimization process. In what follows, we assume that once the investor has seen an overview of the Pareto optimal solutions, he/she takes part of the final solution. Thus, we propose an a-posteriori approach. 

The analyst supporting a-posteriori methodology has to inform the decision maker either providing a list of solutions or providing a visualization of the Pareto front \cite{lotov2008visualizing}. In a tri-objective case, two main approaches have been used to visualize the Pareto frontier: (i) three-dimensional graph, and (ii) decision maps. However,  a new graphical visualization called Level Diagram is proposed in \cite{blasco2008new} to represent n-dimensional Pareto fronts. The Level Diagrams tool, also allows the incorporation of decision makers' preferences and it offers a good tool to help in the decision making process.     

In our proposal, information on preferences is given by the investor, who is willing to achieve a desired aspiration level for each objective function. Let us denote the reference vector for the preferences about green investments defined by the carbon risk score objective function \eqref{eq:tri_03} as $P_{g}$ and the preferences for the loss aversion attitude defined by \eqref{eq:tri_01} as $P_{r}$.  

Concerning the sustainability preferences, we consider three types of green investor profiles. They are defined as follows:

\begin{enumerate}

\item \textit{Weak green investor}. This profile is defined by a low level of aspiration for the carbon risk score $p_{g}^w$.
\item \textit{Moderate green investor}. This profile is defined by a medium level of aspiration for the carbon risk score $p_{g}^m$.
\item \textit{Strong green investor}. This profile is defined by a high level of aspiration for the carbon risk score $p_{g}^s$.

\end{enumerate}

Thus, the reference vector for the green investor could be stated as follows:
\begin{equation}\label{eq:ref_green}
P_{g}=\left[p_{g}^w,\,p_{g}^m,\,p_{g}^s\right]
\end{equation}

Concerning the investor's loss aversion attitude, we consider three types of investor profiles. They are defined as follows:

\begin{enumerate}

\item \textit{Conservative investor}. This profile is characterized by investing in lower-risk securities, namely, a high loss aversion attitude $p_{r}^c$.
\item \textit{Cautious investor}. This profile is defined by a medium risk tolerance, and consequently a moderate loss aversion attitude $p_{r}^k$.
\item \textit{Aggressive investor}. It includes investors that 
actively seek stocks with higher risk—but a chance for higher reward, that is a low loss aversion attitude$p_{r}^a$.
\end{enumerate}

Thus, the reference vector regarding the risk aversion could be stated as follows:
\begin{equation}\label{eq:ref_risk}
P_{r}=\left[p_{r}^c,\,p_{r}^k,\,p_{r}^a\right]
\end{equation}

\begin{figure}[h]
\centering
\parbox[t]{0.475\textwidth}{\centering
	\includegraphics[width = 0.475\textwidth]{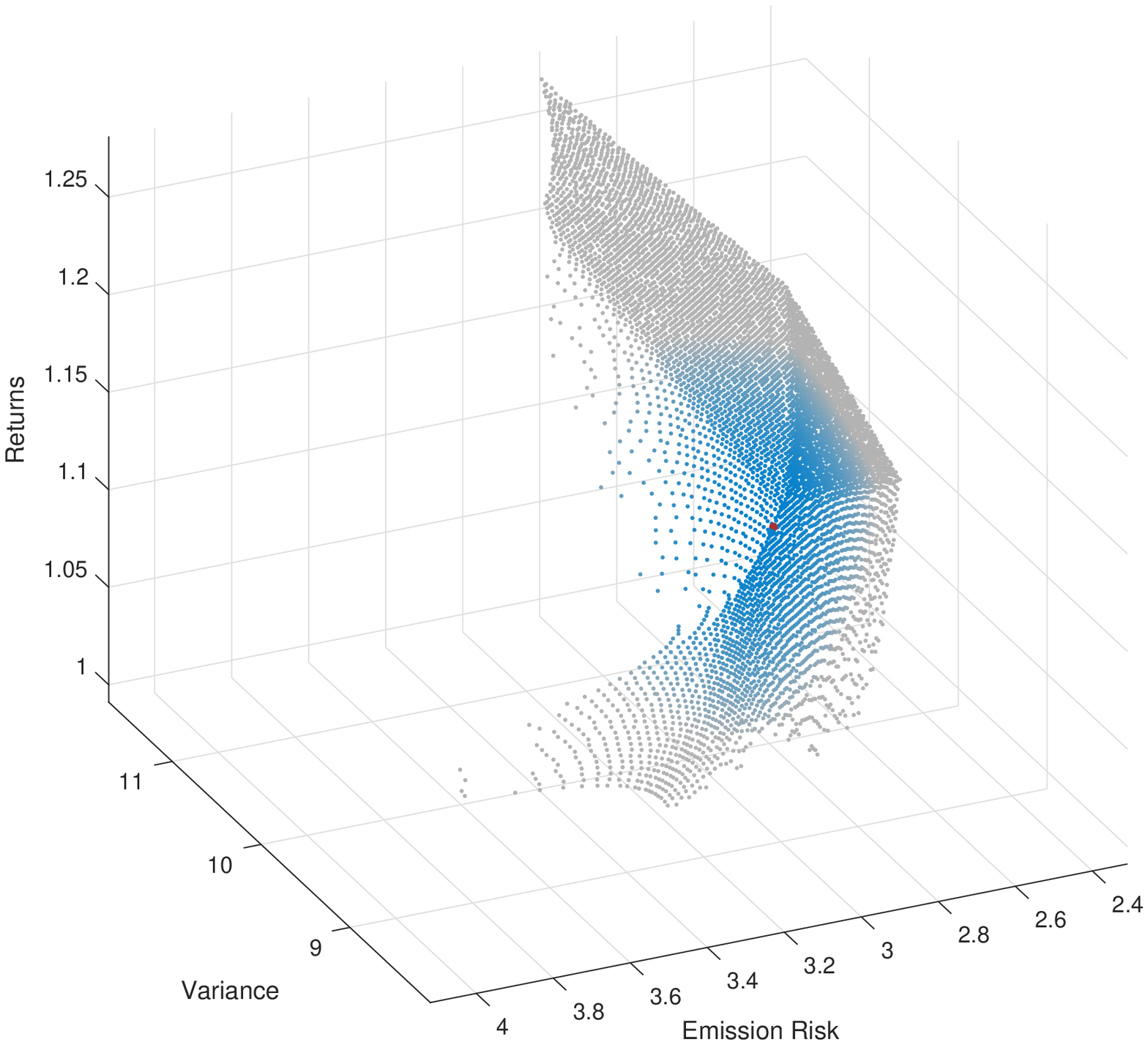}
	\captionof{figure}{$\epsilon$--Pareto front}
	\label{fig:PF}
}
\hfill
\parbox[t]{0.475\textwidth}{\centering
	\includegraphics[width = 0.475\textwidth]{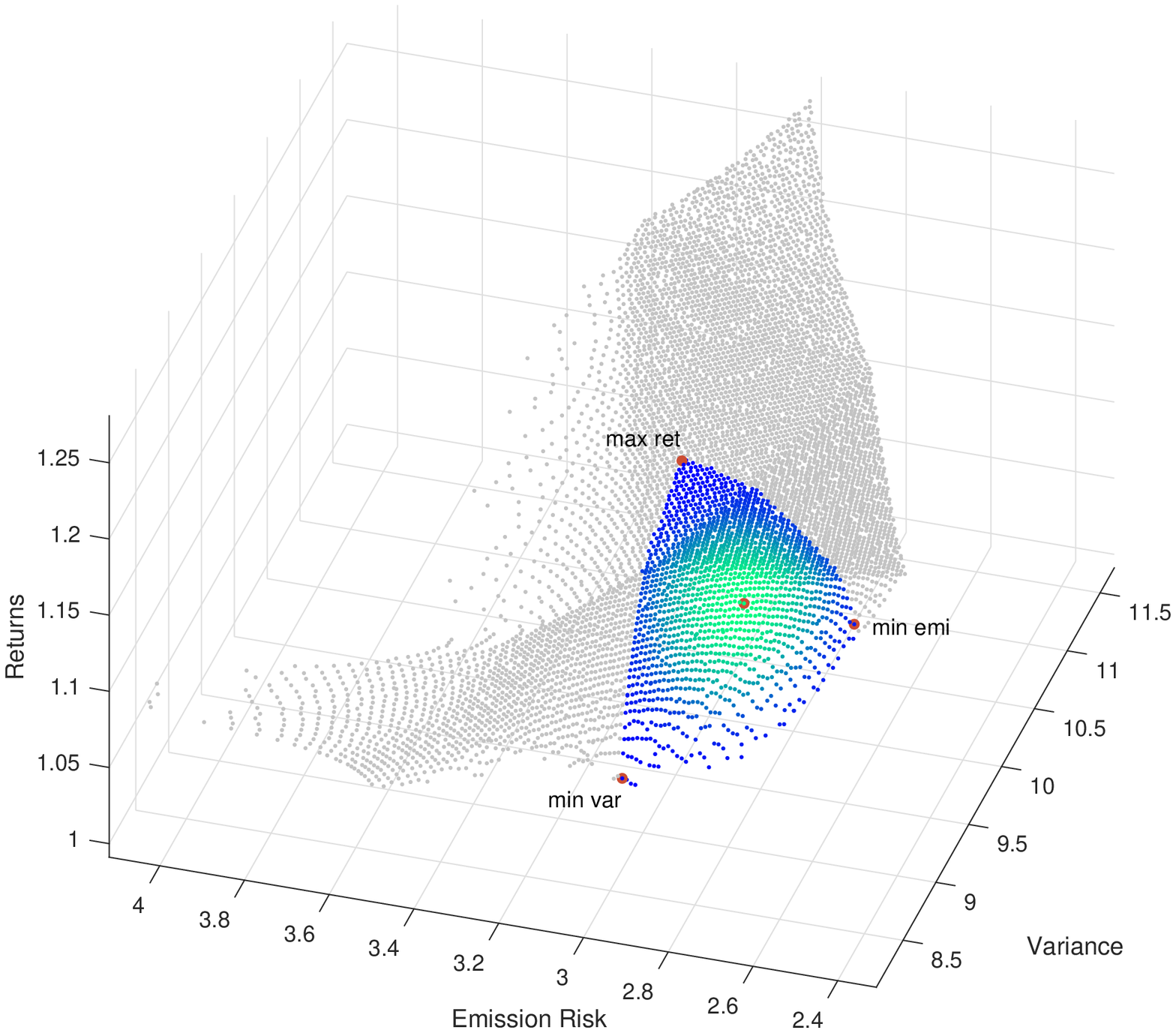}
	\captionof{figure}{$\epsilon$--Pareto front with inverstor's preferences}
	\label{fig:PF_prefs}
}
\end{figure}

\section{Empirical application}\label{s:sexperimetnal}

We use a set of monthly returns on 22 institutional SRI European open-end funds offered in Spain for the period 2009-2019. The empirical information includes the time series of 120 monthly returns and the carbon risk indices. As a previous step the expected return vector $\nu=(\nu_
1,\ldots,\nu_{22})^T$ and the covariance matrix $\Sigma=[\sigma_{ij}]$, ${i,j}=1,\ldots,22$ are computed. For the carbon risk score $c_i$, we use the Morningstar$^{®}$ Portfolio Carbon Risk Score, which indicates the risk that companies face from the transition to a low-carbon economy. In this set, scores $c_i$ range from 0 to 10, where lower scores are better, indicating lower carbon risk levels.  All the numerical information to be used on this opportunity set comes from Morningstar database.

\autoref{tab:parameter-setting} shows the parameter setting applied to the ev-MOGA algorithm. The size of the main population is $Nind_P=10^4$, while the population of the archive $A_k$ is  $Nind_{GA}=500$. For the probability of crossing/mutation we select $P_{m/c}=0.2$. Finally, the space of each objective function has been divided in 300 boxes.

\begin{table}[htbp]
  \centering
  \caption{Parameter setting of the ev-MOGA } 
    \renewcommand{\arraystretch}{1.25}
\begin{tabular}{ll}
    \hline
    Parameter & Value\\
  \hline
    Size of the main population  & $Nind_P=10^4$\\
    Size of the auxiliary population   & $Nind_{GA}=500$ \\
    Maximum algorithm iterations  & $k_{max}=10^5$\\
    Probability of crossing/mutation  & $P_{m/c}=0.2$ \\
    Number of boxes &  300\\
	\hline
\end{tabular}%
  \label{tab:parameter-setting}%
\end{table}

With the aim of analysing the Pareto optimal portfolios retrieved by the ev-MOGA algorithm, we consider the following reference values for each investor profile according the sustainability preferences and the risk aversion.

\newcommand{\percPg}{[25\%, 55\%, 75\%]}
\newcommand{\valoresPg}{[2.803, 3.001, 3.136]}

\newcommand{\percPr}{[50\%, 75\%, 100\%]}
\newcommand{\valoresPr}{[9.575, 10.097, 11.633]}

\begin{itemize}
\item Concerning the sustainability preferences, the green investors are classified in three profiles according to \eqref{eq:ref_green} by using the percentile 25 for the Weak green investor $p_g^w$, 55 for the Moderate green investor $p_g^m$ and 75 for the Strong green investor $p_g^s$. Thus, the reference vector for the green investor yields:
\begin{equation*}
	P_{g}=\percPg=\valoresPg
\end{equation*}


\item Considering the investor loss aversion attitude, the investors are classified in three profiles according to \eqref{eq:ref_risk} by establishing percentiles 50 for a Conservative investor $p_r^c$, 75 for a cautious investor $p_r^k$ and 100 for an Aggressive investor $p_r^a$. Thus, the reference vector regarding the risk aversion becomes:
\begin{equation*}
	P_{r}=\percPr=\valoresPr
\end{equation*}

\end{itemize}

From \autoref{tab:weak}, \autoref{tab:moderate} and \autoref{tab:strong}, a comparison of efficient portfolios for Weak, Moderate and Strong green investors is made in terms of different loss aversion attitude. To this end, for each profile we display a numerical description of the portfolio composition and the objective function values attained by the portfolios. We  highlight in bold optimal funds allocation when achieving the three objectives  simultaneously and we also provide the portfolio weights and the objective values for the strategy involving minimum risk,  minimum carbon risk score and maximum return.

\begin{figure}[h!]
	\centering
	\begin{subfigure}{0.3\textwidth}
		\includegraphics[
			trim = 0cm -1cm 0cm 0cm, clip,
			width = \textwidth]{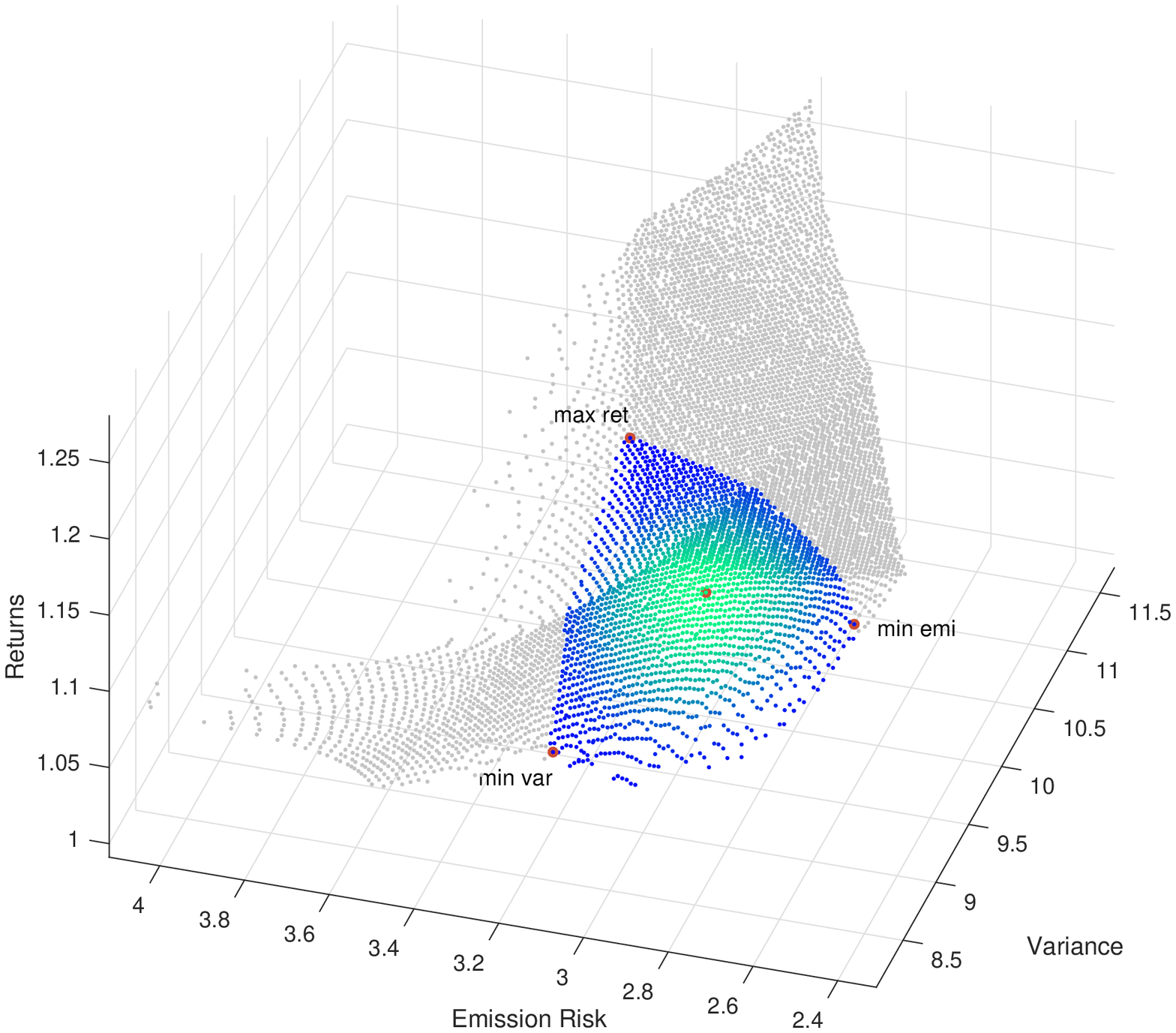}		
		\caption{Conservative investor}
		\label{fig:weak-conservative}			
	\end{subfigure}
	\hspace{0.02\textwidth}
	\begin{subfigure}{0.3\textwidth}
		\includegraphics[
			trim = 0cm -1cm 0cm 0cm, clip,
			width = \textwidth]{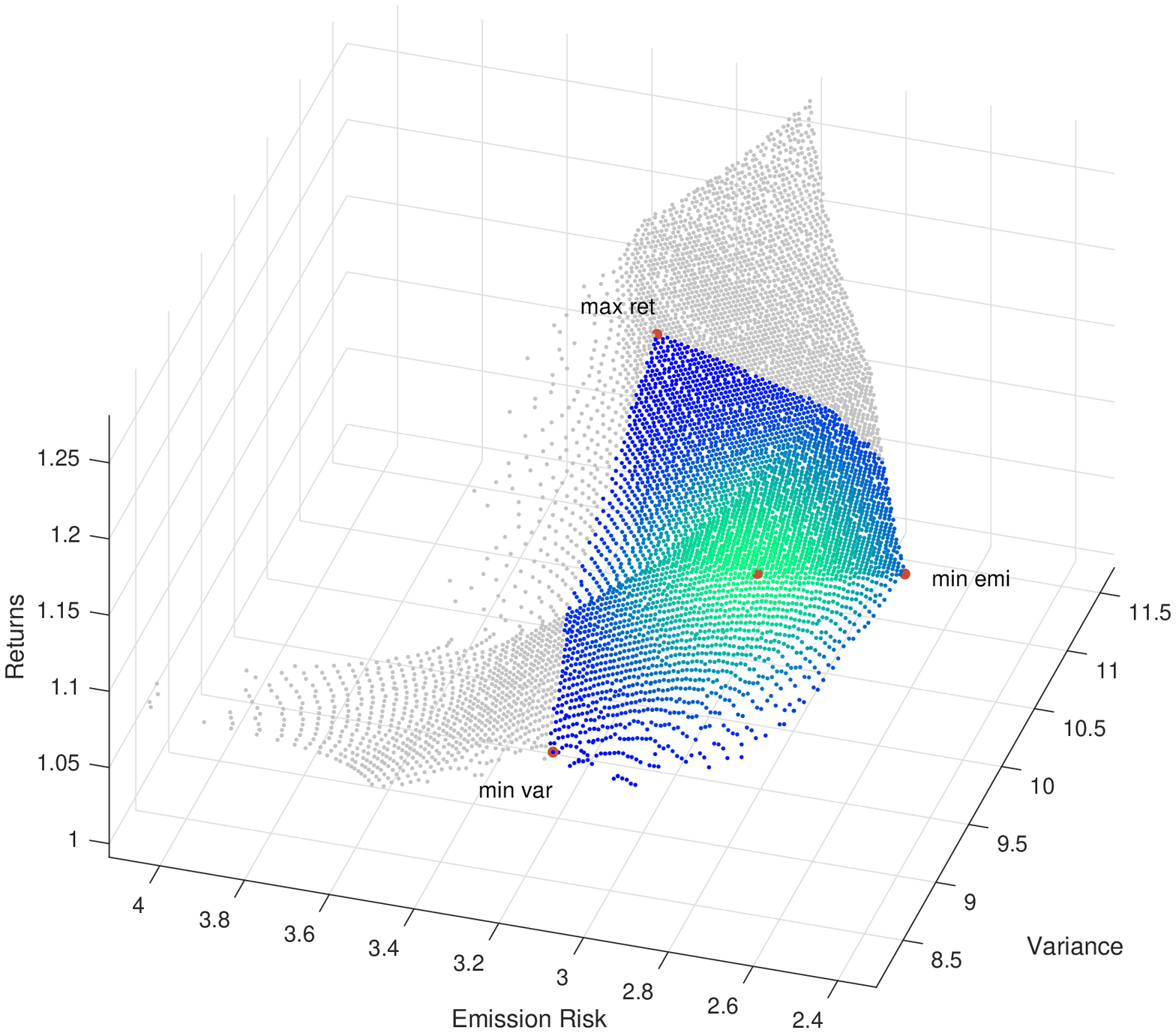}
		\caption{Cautious investor}
		\label{fig:weak-cautious}			
	\end{subfigure}
	\hspace{0.02\textwidth}
	\begin{subfigure}{0.3\textwidth}
		\includegraphics[
			trim = 0cm -1cm 0cm 0cm, clip,
			width = \textwidth]{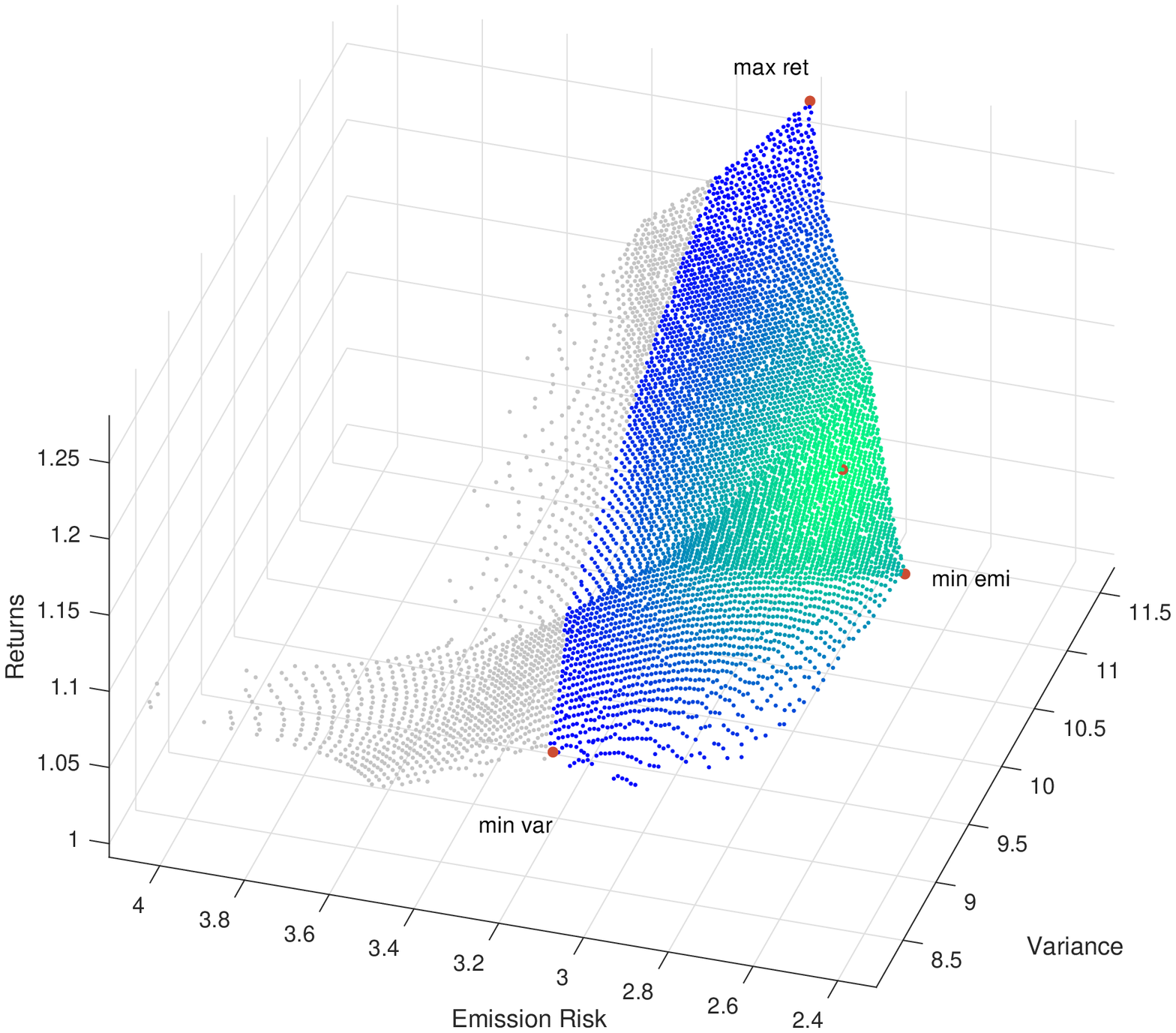}
		\caption{Aggressive investor}
		\label{fig:weak-aggressive}			
	\end{subfigure}
	\bigskip
	\caption{3D Pareto fronts for a Weak green investor profile ($p_g^w$): on the left, for the conservative profile ($p_r^c$); on the centrer for the cautious investor ($p_r^k$); and on the right the aggressive profile ($p_r^a$)}
	\label{fig:weak}
	\bigskip	
\end{figure}

\begin{figure}[h!]
	\centering
	\begin{subfigure}{0.3\textwidth}
		\includegraphics[
			trim = 0cm -1cm 0cm 0cm, clip,
			width = \textwidth]{fig_LC_PF_Moderate_Conservative}		
		\caption{Conservative investor}
		\label{fig:moderate-conservative}			
	\end{subfigure}
	\hspace{0.02\textwidth}
	\begin{subfigure}{0.3\textwidth}
		\includegraphics[
			trim = 0cm -1cm 0cm 0cm, clip,
			width = \textwidth]{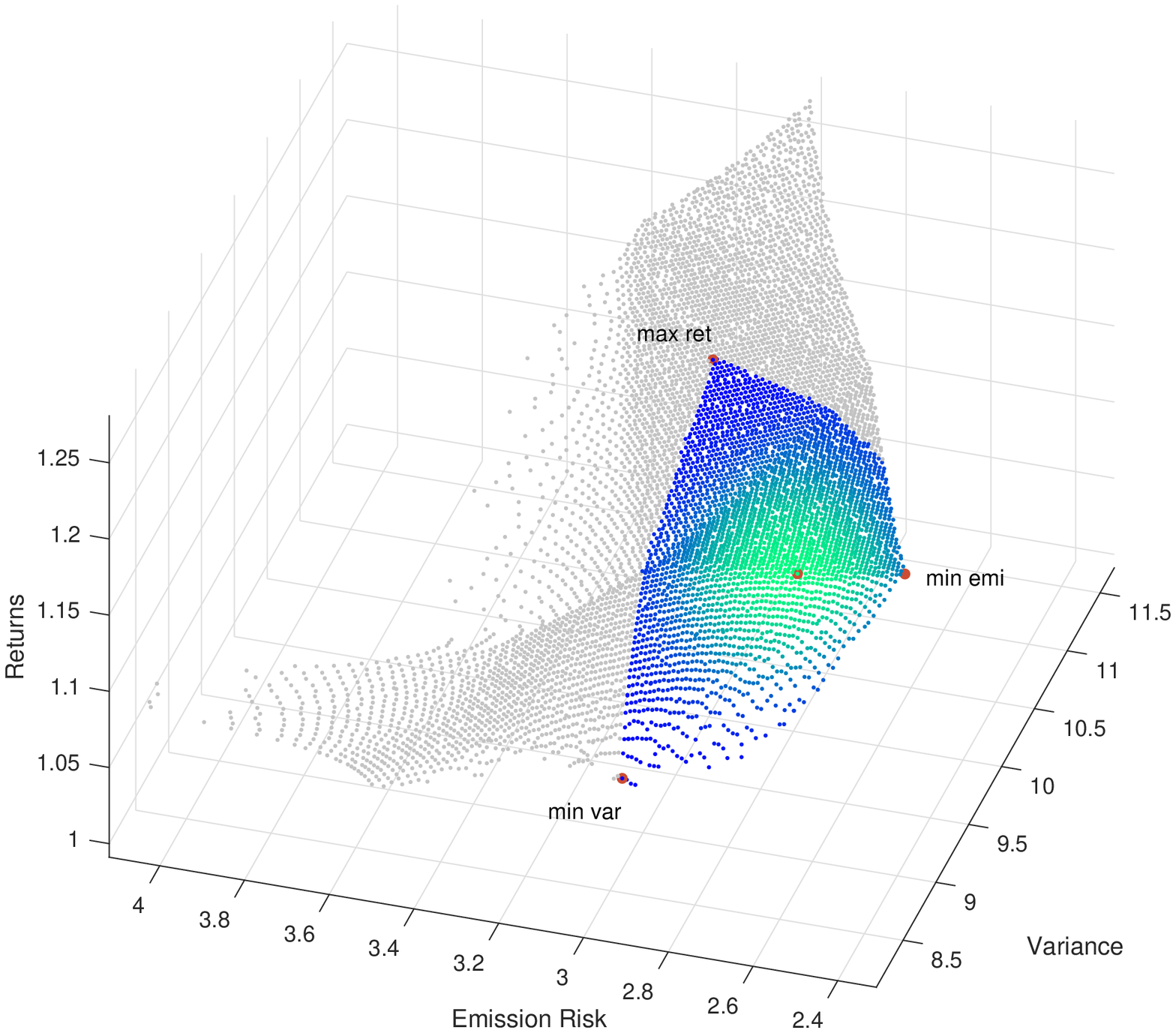}
		\caption{Cautious investor}
		\label{fig:moderate-cautious}			
	\end{subfigure}
	\hspace{0.02\textwidth}
	\begin{subfigure}{0.3\textwidth}
		\includegraphics[
			trim = 0cm -1cm 0cm 0cm, clip,
			width = \textwidth]{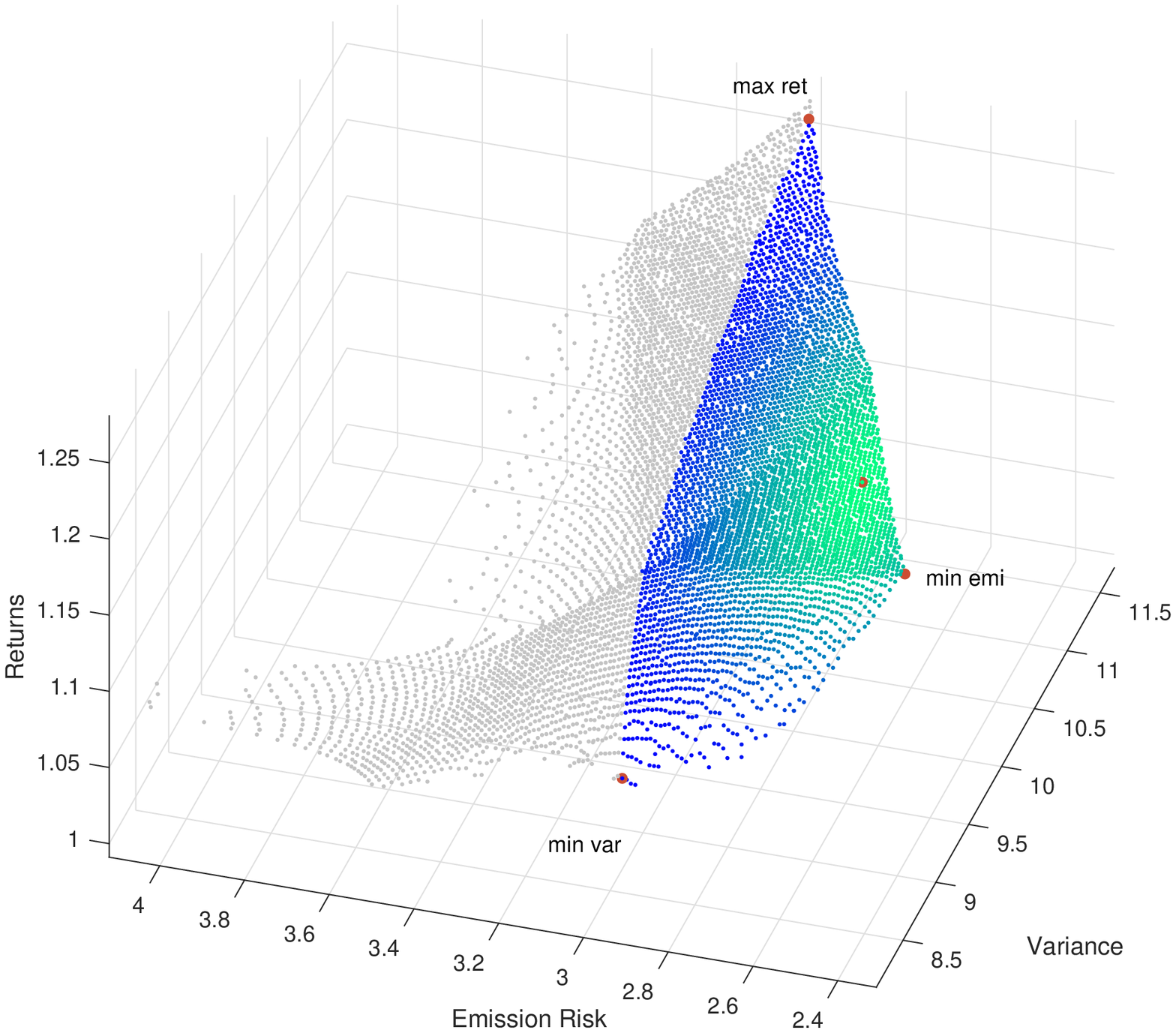}
		\caption{Aggressive investor}
		\label{fig:moderate-aggressive}			
	\end{subfigure}
	\caption{3D Pareto fronts for a Moderate green investor profile. ($p_g^m$): on the left, for the Conservative profile ($p_r^c$); on the centrer for the Cautious investor ($p_r^k$); and on the right for the Aggressive profile ($p_r^a$)}
	\label{fig:moderate}	
\end{figure}

\begin{figure}[h!]
	\centering
	\begin{subfigure}{0.3\textwidth}
		\includegraphics[
			trim = 0cm -1cm 0cm 0cm, clip,
			width = \textwidth]{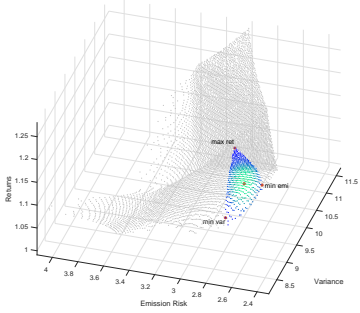}		
		\caption{Conservative investor}
		\label{fig:strong-conservative}			
	\end{subfigure}
	\hspace{0.02\textwidth}
	\begin{subfigure}{0.3\textwidth}
		\includegraphics[
			trim = 0cm -1cm 0cm 0cm, clip,
			width = \textwidth]{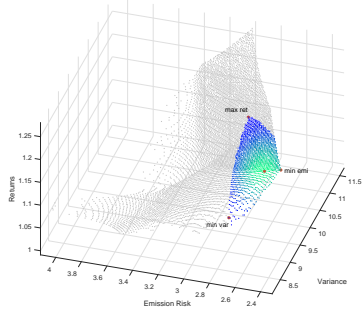}
		\caption{Cautious investor}
		\label{fig:strong-cautious}			
	\end{subfigure}
	\hspace{0.02\textwidth}
	\begin{subfigure}{0.3\textwidth}
		\includegraphics[
			trim = 0cm -1cm 0cm 0cm, clip,
			width = \textwidth]{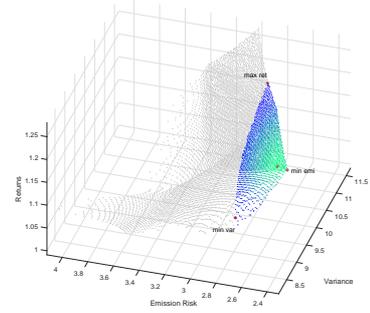}
		\caption{Aggressive investor}
		\label{fig:strong-aggressive}			
	\end{subfigure}
	\caption{3D Pareto fronts for a Strong Green Investor profile ($p_g^s$): on the left, for the conservative profile ($p_r^c$); on the centrer for the cautious investor ($p_r^k$); and on the right for the aggressive profile ($p_r^a$)}
	\label{fig:strong}	
\end{figure}

\autoref{fig:weak}, \autoref{fig:moderate} and \autoref{fig:strong} show the 3D representation of the approximated $\epsilon$-Pareto front, thus providing the non-dominated mean-variance-emission surface for the three types of  Green investor profile and for each level of loss aversion. Notice that, as the level of loss aversion attitude decreases, the Green Investor non-dominated surface (coloured in blue) grows. 

 The results for a Weak green investor profile are displayed in \autoref{tab:weak}. Let us see, for example, the case of an investor's  conservative attitude toward risk.  If the investor wants to  optimize the three objectives simultaneously, the optimal portfolio is given by $F_3$,  $F_{10}$, $F_{11}$, $F_{12}$, , $F_{14}$, $F_{16}$, and $F_{21}$. We can also view the 3D non-dominated surface in \autoref{fig:weak-conservative} in which the whole $\epsilon$-Pareto front is coloured in grey and the investor's region of interest is coloured in blue and green. While the optimum value of the three objectives lies at the centre of the figure, the corner solutions  indicate the optimum objective values involving minimum risk, minimum emissions risk and maximum return. These optimal values are marked by a red dot. Note that the region of interest increases as the investor's risk aversion decreases (see \autoref{fig:weak-cautious} and \autoref{fig:weak-aggressive}).

\newcommand{\espai}{\smallskip}
\newlength{\separaGrupTaula}\setlength{\separaGrupTaula}{2.5em}


\begin{table}[h!]
	\bigskip
	\caption{Weak green investor portfolio composition and objective value function
	}
	\label{tab:weak}
	\centering
	\espai
	\resizebox{\textwidth}{!}{%
	\begin{tabular}{
		*{1}{l} @{\hspace{\separaGrupTaula}} 
		*{8}{r} @{\hspace{\separaGrupTaula}} 
		c c c @{\hspace{\separaGrupTaula}}
		l}\toprule
		Risk profile	&	$F_{3}$	&	$F_{10}$	&	$F_{11}$	&	$F_{12}$	&	$F_{13}$	&	$F_{14}$	&	$F_{16}$	&	$F_{21}$	&	Risk	&	Ret.	&	Emiss\\\midrule\\[-2ex]

\multirow{4}{*}{Conservative}	&	\textbf{20.0}	&	\textbf{20.0}	&	\textbf{20.0}	&	\textbf{13.1}	&	\textbf{0.0}	&	\textbf{6.1}	&	\textbf{19.0}	&	\textbf{1.8}	&	\textbf{9.122}	&	\textbf{1.145}	&	\textbf{2.871}	&	\textbf{opt}\\[0ex]

	&	20.0	&	20.0	&	19.6	&	2.2	&	0.0	&	1.6	&	16.7	&	19.9	&	8.462	&	1.079	&	3.127	&	min var\\[0ex]

	&	20.0	&	20.0	&	17.3	&	0.0	&	0.0	&	19.0	&	19.4	&	4.3	&	9.559	&	1.104	&	2.591	&	min emi\\[0ex]

	&	20.0	&	13.7	&	20.0	&	20.0	&	8.9	&	0.0	&	17.4	&	0.0	&	9.565	&	1.202	&	3.118	&	max ret\\[0ex]

\\[-1ex]
\multirow{4}{*}{Cautious}	&	\textbf{20.0}	&	\textbf{20.0}	&	\textbf{20.0}	&	\textbf{12.0}	&	\textbf{0.0}	&	\textbf{9.9}	&	\textbf{18.1}	&	\textbf{0.0}	&	\textbf{9.312}	&	\textbf{1.148}	&	\textbf{2.778}	&	\textbf{opt}\\[0ex]

	&	20.0	&	20.0	&	19.6	&	2.2	&	0.0	&	1.6	&	16.7	&	19.9	&	8.462	&	1.079	&	3.127	&	min var\\[0ex]

	&	20.0	&	20.0	&	20.0	&	0.0	&	0.0	&	20.0	&	20.0	&	0.0	&	9.801	&	1.123	&	2.506	&	min emi\\[0ex]

	&	20.0	&	5.6	&	20.0	&	20.0	&	13.1	&	1.3	&	20.0	&	0.0	&	10.086	&	1.230	&	3.136	&	max ret\\[0ex]

\\[-1ex]
\multirow{4}{*}{Aggressive}	&	\textbf{20.0}	&	\textbf{10.3}	&	\textbf{20.0}	&	\textbf{14.7}	&	\textbf{0.0}	&	\textbf{18.4}	&	\textbf{16.6}	&	\textbf{0.0}	&	\textbf{9.923}	&	\textbf{1.175}	&	\textbf{2.671}	&	\textbf{opt}\\[0ex]

	&	20.0	&	20.0	&	19.6	&	2.2	&	0.0	&	1.6	&	16.7	&	19.9	&	8.462	&	1.079	&	3.127	&	min var\\[0ex]

	&	20.0	&	20.0	&	20.0	&	0.0	&	0.0	&	20.0	&	20.0	&	0.0	&	9.801	&	1.123	&	2.506	&	min emi\\[0ex]

	&	20.0	&	0.0	&	20.0	&	20.0	&	20.0	&	20.0	&	0.0	&	0.0	&	11.633	&	1.271	&	3.014	&	max ret\\[0ex]

\bottomrule

	\end{tabular}
	}
\end{table}


\begin{table}[h!]
	\bigskip
	\caption{Moderate green investor portfolio composition and objective value function
	}
	\label{tab:moderate}
	\centering
	\espai
	\resizebox{\textwidth}{!}{%
	\begin{tabular}{
		*{1}{l} @{\hspace{\separaGrupTaula}} 
		*{8}{r} @{\hspace{\separaGrupTaula}} 
		c c c @{\hspace{\separaGrupTaula}}
		l}\toprule
		Risk profile	&	$F_{3}$	&	$F_{10}$	&	$F_{11}$	&	$F_{12}$	&	$F_{13}$	&	$F_{14}$	&	$F_{16}$	&	$F_{21}$	&	Risk	&	Ret.	&	Emiss\\\midrule\\[-2ex]

\multirow{4}{*}{Conservative}	&	\textbf{20.0}	&	\textbf{20.0}	&	\textbf{20.0}	&	\textbf{8.7}	&	\textbf{0.0}	&	\textbf{9.3}	&	\textbf{19.6}	&	\textbf{2.4}	&	\textbf{9.231}	&	\textbf{1.135}	&	\textbf{2.788}	&	\textbf{opt}\\[0ex]

	&	20.0	&	20.0	&	13.3	&	1.1	&	0.0	&	9.7	&	17.9	&	18.0	&	8.721	&	1.061	&	2.962	&	min var\\[0ex]

	&	20.0	&	20.0	&	17.3	&	0.0	&	0.0	&	19.0	&	19.4	&	4.3	&	9.559	&	1.104	&	2.591	&	min emi\\[0ex]

	&	20.0	&	11.3	&	20.0	&	20.0	&	4.0	&	4.7	&	20.0	&	0.0	&	9.562	&	1.193	&	2.970	&	max ret\\[0ex]

\\[-1ex]
\multirow{4}{*}{Cautious}	&	\textbf{20.0}	&	\textbf{20.0}	&	\textbf{20.0}	&	\textbf{8.0}	&	\textbf{0.0}	&	\textbf{12.5}	&	\textbf{19.5}	&	\textbf{0.0}	&	\textbf{9.436}	&	\textbf{1.139}	&	\textbf{2.696}	&	\textbf{opt}\\[0ex]

	&	20.0	&	20.0	&	13.3	&	1.1	&	0.0	&	9.7	&	17.9	&	18.0	&	8.721	&	1.061	&	2.962	&	min var\\[0ex]

	&	20.0	&	20.0	&	20.0	&	0.0	&	0.0	&	20.0	&	20.0	&	0.0	&	9.801	&	1.123	&	2.506	&	min emi\\[0ex]

	&	20.0	&	4.5	&	20.0	&	20.0	&	7.8	&	7.7	&	20.0	&	0.0	&	10.077	&	1.218	&	2.968	&	max ret\\[0ex]

\\[-1ex]
\multirow{4}{*}{Aggressive}	&	\textbf{20.0}	&	\textbf{13.1}	&	\textbf{20.0}	&	\textbf{10.6}	&	\textbf{0.0}	&	\textbf{20.0}	&	\textbf{16.3}	&	\textbf{0.0}	&	\textbf{9.919}	&	\textbf{1.161}	&	\textbf{2.612}	&	\textbf{opt}\\[0ex]

	&	20.0	&	20.0	&	13.3	&	1.1	&	0.0	&	9.7	&	17.9	&	18.0	&	8.721	&	1.061	&	2.962	&	min var\\[0ex]

	&	20.0	&	20.0	&	20.0	&	0.0	&	0.0	&	20.0	&	20.0	&	0.0	&	9.801	&	1.123	&	2.506	&	min emi\\[0ex]

	&	20.0	&	0.0	&	20.0	&	20.0	&	17.1	&	19.7	&	3.2	&	0.0	&	11.380	&	1.261	&	2.968	&	max ret\\[0ex]

\bottomrule

	\end{tabular}
	}
\end{table}

In \autoref{tab:moderate} and \autoref{tab:strong}, the optimal portfolio and the values of the objective functions are shown for the corresponding investor profiles. As expected, the minimum region of interest correspond to an Strong green investor with a conservative attitude toward risk as shown in \autoref{fig:strong-conservative}. 

\clearpage

\begin{table}[h!]
	\bigskip
	\caption{Strong green investor portfolio composition and objective value function
	}
	\label{tab:strong}
	\centering
	\espai
	\resizebox{\textwidth}{!}{%
	\begin{tabular}{
		*{1}{l} @{\hspace{\separaGrupTaula}} 
		*{8}{r} @{\hspace{\separaGrupTaula}} 
		c c c @{\hspace{\separaGrupTaula}}
		l}\toprule
		Risk profile	&	$F_{3}$	&	$F_{10}$	&	$F_{11}$	&	$F_{12}$	&	$F_{13}$	&	$F_{14}$	&	$F_{16}$	&	$F_{21}$	&	Risk	&	Ret.	&	Emiss\\\midrule\\[-2ex]

\multirow{4}{*}{Conservative}	&	\textbf{20.0}	&	\textbf{20.0}	&	\textbf{20.0}	&	\textbf{2.6}	&	\textbf{0.0}	&	\textbf{13.3}	&	\textbf{19.9}	&	\textbf{4.2}	&	\textbf{9.353}	&	\textbf{1.118}	&	\textbf{2.698}	&	\textbf{opt}\\[0ex]

	&	20.0	&	19.9	&	10.9	&	0.6	&	0.0	&	15.5	&	19.5	&	13.6	&	9.054	&	1.062	&	2.795	&	min var\\[0ex]

	&	20.0	&	20.0	&	17.3	&	0.0	&	0.0	&	19.0	&	19.4	&	4.3	&	9.559	&	1.104	&	2.591	&	min emi\\[0ex]

	&	20.0	&	13.4	&	20.0	&	18.4	&	0.0	&	12.3	&	15.9	&	0.0	&	9.558	&	1.176	&	2.804	&	max ret\\[0ex]

\\[-1ex]
\multirow{4}{*}{Cautious}	&	\textbf{20.0}	&	\textbf{20.0}	&	\textbf{20.0}	&	\textbf{4.0}	&	\textbf{0.0}	&	\textbf{16.0}	&	\textbf{20.0}	&	\textbf{0.0}	&	\textbf{9.601}	&	\textbf{1.131}	&	\textbf{2.604}	&	\textbf{opt}\\[0ex]

	&	20.0	&	19.9	&	10.9	&	0.6	&	0.0	&	15.5	&	19.5	&	13.6	&	9.054	&	1.062	&	2.795	&	min var\\[0ex]

	&	20.0	&	20.0	&	20.0	&	0.0	&	0.0	&	20.0	&	20.0	&	0.0	&	9.801	&	1.123	&	2.506	&	min emi\\[0ex]

	&	20.0	&	3.8	&	20.0	&	20.0	&	2.4	&	13.8	&	20.0	&	0.0	&	10.070	&	1.204	&	2.803	&	max ret\\[0ex]

\\[-1ex]
\multirow{4}{*}{Aggressive}	&	\textbf{20.0}	&	\textbf{20.0}	&	\textbf{20.0}	&	\textbf{4.6}	&	\textbf{0.0}	&	\textbf{19.5}	&	\textbf{15.8}	&	\textbf{0.0}	&	\textbf{9.695}	&	\textbf{1.133}	&	\textbf{2.569}	&	\textbf{opt}\\[0ex]

	&	20.0	&	19.9	&	10.9	&	0.6	&	0.0	&	15.5	&	19.5	&	13.6	&	9.054	&	1.062	&	2.795	&	min var\\[0ex]

	&	20.0	&	20.0	&	20.0	&	0.0	&	0.0	&	20.0	&	20.0	&	0.0	&	9.801	&	1.123	&	2.506	&	min emi\\[0ex]

	&	20.0	&	0.0	&	20.0	&	20.0	&	7.3	&	19.6	&	13.1	&	0.0	&	10.716	&	1.229	&	2.803	&	max ret\\[0ex]

\bottomrule

	\end{tabular} 
	}
\end{table}

\section{Conclusions}

In this paper, we have proposed a new application of the ev-MOGA algorithm to handle a tri-criterion portfolio optimization problem in which the third criterion is the carbon risk score of the portfolio. Moreover, we have incorporated the investor's preferences regarding the risk emissions and the loss aversion attitude into the solution process by defining different investor profiles. This allows us to propose a solution to the investor in terms of their sustainability and risk preferences.  

Given the urgency around climate change, investors are becoming increasingly aware of the need to make the transition to a lower carbon economy and to address climate change  related risk. It is claimed that new methodological tools are needed to help investors to align themselves with the preservation of the planet without compromising returns and to more thoughtfully consider carbon risk in the investment decision making framework. In recent years, some rating agencies have introduced eco-labels for mutual funds which allow for the measurement of the risk that  companies face in the transition to a low carbon economy. We have reviewed the literature addressing the extended M-V portfolio optimization problem. While some scholars have developed exact methodologies to derive the non-dominated surface, in recent years there is an increase number of contributions applying heuristic methodologies such as Multi-objective Evolutionary Algorithms. Thus, we implement a tri-criterion portfolio optimization problem for returns, risk and emission risks by means of an heuristic methodology based on the concept of $\epsilon$- dominance called ev-MOGA. To the best of our knowledge this is the first time that this methodology is used to derive the non-dominated surface in three dimensions including mean, variance and carbon-related risks. The ev-MOGA allows us to obtain a 3D-Pareto front in a well-distributed manner with limited memory resources.

To better understand the trade-off between the three objectives we have introduced an a-posteriori approach to include the investors' preferences about green investments and risk aversion. So, by considering different investor profiles we can provide a more approximated solution to the investor according to their preferences. Because of the possibility to obtain an efficient frontier in three dimensions while including the preferences on risk and sustainability, we believe this is a useful tool for investors, especially for those who are willing to rebalance their portfolios towards more climate-conscious firms.

Finally, as the mutual fund industry is an ideal setting to test our approach, we have used a set of institutional SRI European open-end funds for illustrative purposes. In the numerical experiments we have analyzed the portfolio generated according to the investor profiles. The results obtained show that the region of interest increases as the investor's risk aversion decreases, namely, aggressive investors looking for high returns are allowed to invest in funds with a lower level of carbon risk scores. Thus, we conclude that green investors have a leeway to decrease the emission risk of the portfolio at even no cost to risk and returns.



\bibliographystyle{abbrv}
\bibliography{Low_Carbon}

\end{document}